\documentclass[12pt]{iopart}
\usepackage{graphicx, algorithm, algorithmic, hyperref}
\begin{document}

\title[Titan from \emph{Kepler K2}]{\emph{k}-Means Aperture Optimization Applied to \emph{Kepler K2} Time Series Photometry of Titan}

\author{Alex H. Parker$^{1*}$, Sarah M. H{\"o}rst$^{2,3}$, Erin L. Ryan$^{4}$, Carly J. A. Howett$^{1}$}

\address{$^1$: Southwest Research Institute, Boulder, Colorado, 80302. $^{2}$: Johns Hopkins University Department of Earth and Planetary Science, Baltimore, MD, 21218. $^{3}$: Space Telescope Science Institute, Baltimore, MD, 21218. $^{4}$: SETI Institute, Mountain View, CA, 94043.}
\ead{aparker@boulder.swri.edu}
\begin{abstract}

Motivated by the \emph{Kepler K2} time series of Titan, we present an aperture optimization technique for extracting photometry of saturated moving targets with high temporally- and spatially-varying backgrounds. Our approach uses $k$-means clustering to identify interleaved families of images with similar Point-Spread Function and saturation properties, optimizes apertures for each family independently, then merges the time series through a normalization procedure. By applying $k$-means aperture optimization to the \emph{K2} Titan data, we achieve $\leq0.33\%$ photometric scatter in spite of background levels varying from 15\% to 60\% of the target's flux. We find no compelling evidence for signals attributable to atmospheric variation on the timescales sampled by these observations. We explore other potential applications of the $k$-means aperture optimization technique, including testing its performance on a saturated \emph{K2} eclipsing binary star. We conclude with a discussion of the potential for future continuous high-precision photometry campaigns for revealing the dynamical properties of Titan's atmosphere. 

\end{abstract}

\submitto{Publication of the Astronomical Society of the Pacific}

\section{Introduction}

Saturn's largest moon Titan has a cold, dense ($\sim$1.5 bar, 94 K) N$_{2}$ ($\sim$98\%) and CH$_{4}$ ($\sim$2\%) atmosphere that is rendered nearly opaque by photochemically-generated organic aerosols composed of hydrocarbons and nitriles (e.g., West et al. 2014). Titan's atmosphere drives the only known extant non-terrestrial hydrological cycle, including clouds, rain, fluvial channels, and lakes/seas (see e.g., H\"orst 2017). The precipitable column of methane in Titan's atmosphere is far larger than that of water in the Earth's atmosphere (Atreya et al. 2006). Titan's atmospheric dynamics and chemistry are connected and complex: the atmosphere is highly stratified (Strobel et al. 2009), at altitude it rotates much faster than the moon's surface (e.g., Lebonnois et al. 2014), it supports both short- and long-lived storm systems (e.g., Griffith et al. 2014), and some of its molecular constituents result from exogenic sources (H\"{o}rst et al. 2008). From a photochemical standpoint, Titan's atmosphere is a frozen analog to the early Earth's, and understanding the processes that govern its atmosphere are of great astrobiological interest. For more information about Titan's atmosphere, see the recent review by H\"orst (2017). 

Earth-based monitoring programs have used variability of Titan's brightness in and out of a methane spectral absorption band to identify the onset of storms on Titan and trigger subsequent follow-up from large observing facilities (see e.g., Bouchez and Brown, 2005). In late 2016, the \emph{Kepler} Space Telescope conducted a similar monitoring campaign of Titan during its \emph{K2} mission. This campaign lacked the spectral information of previous efforts, but had the potential to deliver substantially higher photometric precision and better sampling of dynamically-relevant timescales. In 2001, transient clouds were witnessed daily over a 16-night Palomar observing program, and they varied on timescales as short as 3 hours (Bouchez and Brown, 2005), a timescale very well sampled by the continuous  \emph{K2} Long Cadence (LC) 29.4 minute image cadence.

In the following sections, we describe the \emph{K2} observations of Titan and the challenges inherent in analyzing them, describe our methods for modeling scattered light from Saturn and for optimizing photometric apertures for extracting the Titan lightcurve, and present the extracted lightcurve. We conclude with a brief discussion of these results and the potential applications of our techniques to existing and future datasets, including a demonstration application on K2 data of a saturated eclipsing binary star. A repository containing a \emph{Python} implementation of the $k$-means aperture optimization algorithms as described in this paper is available online\footnote{ \url{https://github.com/alex-parker/kmao}}.

\section{{\it K2} Observations}

Image data from  \emph{K2} are delivered as 1D pixel masks and collections of these masks must be assembled to reconstruct a 2D image. We assemble the Titan {\it K2} image dataset from the delivered pixel masks using the same approach as Ryan et al. (2017). An example image from the sequence is shown in Figure 1.

The Titan \emph{K2} dataset includes more than 400 Long Cadence (LC) images of the field that Titan and Saturn move through but which precede the arrival of Titan and Saturn into the field. We leverage this sample of sky-only data to define stable World Coordinate System (WCS) parameters for the images of interest using the following procedure. For each image in the sequence containing Titan, we find the image from the preceding sequence \emph{not} containing Titan that minimizes the sum of the squares of the difference for pixels that (1) do not contain substantial Titan or Saturn flux, and (2) have been high-pass filtered to remove low-frequency sky variations.  We use a list of eight stars linked to the STScI GSC 2.3 (see Table 1) to determine WCS parameters for the matched comparison image, and apply these parameters to the Titan/Saturn image of interest. Finally, we subtract the matched stars-only image from the Titan/Saturn image to remove starlight contamination.

All told, 182 images were identified which contained Titan sufficiently far from the chip or mask boundary and sufficiently far from the saturation effects of Saturn to be considered for further photometric analysis. These images span the UTC range 2016-Dec-03 17:37:31.0 to 2016-Dec-07 10:52:45.3.

\begin{table}
\begin{center}
\begin{tabular}{|lcc|}

\multicolumn{3}{c}{{\bf Table 1:} WCS Reference Stars} \\
\hline
gsc2ID& RA (J2000)& Dec (J2000) \\
\hline
S223013276& 252.663855$^\circ$& -21.102296$^\circ$ \\
S223013285185& 252.674811$^\circ$& -21.089284$^\circ$  \\
S223013244& 252.604796$^\circ$& -21.073618$^\circ$  \\
S223013285789& 252.514405$^\circ$& -21.083814$^\circ$  \\
S223013282461& 252.715532$^\circ$& -21.114431$^\circ$  \\
S223013284635& 252.744156$^\circ$& -21.094139$^\circ$  \\
S223013285789& 252.514405$^\circ$& -21.083814$^\circ$ \\
S223013287933& 252.464452$^\circ$& -21.063552$^\circ$ \\
\hline
\end{tabular}
\end{center}
\end{table}

\section{Saturn Scattered Light}

The chief challenge of extracting accurate photometry of Titan is removing the contribution from Saturn's scattered light. We tested several models of the scattered light, including constructing a PSF model and scaling it to fit to data in an annulus around Titan, building a 3D (Saturn-relative sky plane position and time) flux interpolant from which to predict flux at any point in a given Titan aperture at a given time, and others. The selected model performed far better and more stably than these other methods without introducing substantial model complexity. 

We first translate the image (x,y) pixel coordinates into a scaled polar coordinate system centered on Saturn's location in the focal plane. The coordinates are defined as 

\begin{equation*}
\eqalign{
r_{s} = \sqrt{(x-x_{s})^2 + (y-y_{s})^2}\cr
\theta_{s} = s \times r_{s} \times \arctan_{2}(y-y_{s}, x-x_{s}),
}
\label{eq1} 
\end{equation*}

\noindent where $s$ is a scaling factor designed to prioritize either the radial behavior or azimuthal behavior in the scattered light profiles. We determined by brute force that $s=2$ was optimal for the portion of the Saturn scattered light field that Titan passed through in these observations, weakly prioritizing radial behavior. 

We then convert the raw \emph{Kepler} flux into $log_{10}$(flux), and build a thin-plate Radial Basis Function interpolant (using \emph{scipy.interpolate.Rbf} with an $r^2 \log(r)$ basis function; c.f. Millman \& Aivazis 2011) in the Saturn-radial coordinate system and interpolating $\log_{10}$(flux). This interpolant is built on data drawn from a 2-pixel wide boundary around any proposed Titan aperture, and we then use it to predict $log_{10}$(flux) across that aperture. 

Of all tested methods, this approach produced the lowest scatter for Titan's flux extracted from any given static aperture, the lowest scatter after $k$-means aperture optimization (see following section), and the lowest Spearman's rank correlation coefficient between the predicted sky values in the aperture and the extracted Titan lightcurve. Finally, visual inspection of the predicted scattered light field over the photometric aperture showed smooth boundaries and qualitatively accurate reproduction of discrete features (e.g., diffraction spikes) of the Saturn PSF as they swept across the aperture. 

Since this approach predicts the flux inside an aperture based on an extrapolation of the flux observed outside the aperture, it grows less accurate the larger the aperture is. This drives us to a preference for as small an aperture as possible, and the selection of this aperture is described in the following section.


\section{$K$-Means Aperture Optimization}

Defining the optimum photometry aperture for a \emph{Kepler} time series is critical for achieving high precision measurements, and substantial effort has gone into determining these for stationary targets (e.g., Smith et al. 2016). For a moving, saturated target, however, a single aperture must be made quite large to encompass a large fraction of the PSF and the saturation charge bleed spikes. Depending on the sub-pixel position and integrated motion of the target over a 30 minute LC exposure, the shape and extent of the charge bleed spikes can vary substantially. Pixel-Level Decorrelation (PLD; Luger et al. 2016, Luger et al. 2018) can deliver high precision on saturated, relatively stationary targets by incorporating the PLD vectors from neighboring stars to recover information lost due to saturation, but this technique leverages the relative stability of a network of stationary sources has not been generalized to moving sources. Sky plane motion decorrelates pixel values from one frame to the next independently of the spacecraft pointing. To circumvent this, other \emph{K2} studies of bright solar system targets have adopted extremely large moving apertures to capture as much of the PSF as possible (e.g., Neptune by Simon et al. 2016). For Titan, adopting a single large aperture would result in substantial contamination from Saturn's time-varying PSF. Instead, we would like to construct an aperture that is \emph{as small as possible} on each image to limit the contribution of Saturn scattered light. However, without ultra-accurate PSF models, a frame-by-frame varying aperture would be prohibitively difficult to calibrate. The methods presented here are a hybrid of these two end-member approaches that results in a workable compromise between aperture losses and background contamination.

We optimize a small \emph{set} of $k$ apertures, each applied to a unique subset of images with similar properties, which enable the use of compact apertures by enabling internal calibration of the aperture correction for $k-1$ of the apertures with respect to a single reference aperture. This allows different apertures to be applied to images where, for example, the target falls near the center of a pixel versus images where it falls near a pixel boundary, enabling separate optimization of the apertures applied to the differing resultant charge bleed spikes, PSF sampling, and so forth. These independent apertures also naturally detrend pointing-based systematics, as pointing-related processes express themselves in detectable ways on the image plane during a single exposure. We define these apertures by applying $k$-means clustering on a feature set derived from the target image at each epoch. $K$-means clustering assigns membership of a set of $n$ samples of into $k$ groups of equal variance in an $N$-dimensional feature space, where each sample is assigned membership to the cluster with the nearest mean in that feature space (see, e.g., MacQueen 1967). In our case, we extract an ordered 1D list of pixel values from the image of a target drawn from a constant large aperture on the focal plane, normalize this list to its mean, and consider this normalized list to be our feature set for performing $k$-means clustering. This process creates $k$ clusters of images with similar properties -- effectively, similar motion blur, saturation, and sub-pixel target positioning, recovered directly from the images rather than from a prediction. Because the assignment of an image to a cluster is determined only by its normalized pixel values, clusters are populated by images that are relatively uniformly sampled from throughout the time series. This property of interleaved sampling enables the use of a finite-difference based scheme to robustly calibrate the aperture corrections for $k-1$ of the apertures with respect to a reference aperture. The full process for defining these apertures follows:

\begin{enumerate}

\item Determine accurate WCS parameters for each image and subtract an image of the background starfield.

\item Using the updated star-based WCS parameters and a prediction of the target's equatorial position as seen from  \emph{K2} in each frame, co-register the target images to the nearest integer pixel. Determine and subtract a preliminary scattered light model (e.g., as described in Section 3) from each image.

\item For each of the $n$ images, flatten the array of $N$ pixel values from a large aperture around the target into a 1D list of $N$ features. Stack these feature arrays into a 2D ($n \times N$) array of features extracted from the image set.

\item Perform $k$-means clustering on this 2D feature array to identify $k$ sub-groups of images with similar properties.

\item Stack each cluster of similar target images to generate a cluster average image. Trim pixels from an initial large aperture by iteratively removing the pixel with the lowest estimated signal-to-noise ratio until the aperture contains 99\% of the total target flux estimated from the original large aperture. Dilate this trimmed aperture by one pixel.

\item Extract target photometry using the dilated, trimmed apertures applied to the clusters of target images from which each was derived. 

\end{enumerate}

This results in $k$ independent lightcurves sampled at epochs that are interleaved in time. To merge them into a single record, we identify the cluster with the largest number of members and define it as our reference lightcurve. We then determine and apply multiplicative aperture corrections for each other cluster in order to bring the full set into agreement with itself by minimizing the point-to-point scatter $\sigma_p$, defined as:

\begin{equation*}
\eqalign{
\sigma_{p} = 0.8166 \times \sqrt{\frac{\sum_{i=2}^{n-1}(d - \bar{d})^2}{n-1}}\mbox{ ,   where:}\cr 
d = (y_{i} - \frac{1}{2}(y_{i-1}+y_{i+1})
}
\label{eq2} 
\end{equation*}

\noindent This is a scaled L2 norm on the second-order finite difference. Intuitively, this represents a high-frequency estimate of the standard deviation of measurements around an unknown underlying signal. This formulation assumes a continuous, uniformly-sampled sequence that is sorted in the  independent variable (time, in our case). The leading factor 0.8166 scales $\sigma_{p}$ to match the standard deviation of a sequence if the underlying signal is constant in time and the data are normally distributed; this scaling factor is included merely for convenience of comparing to other estimates of intrinsic scatter. We adopt $\sigma_{p}$ as our objective function to for minimization to determine aperture corrections, similar to the use of Total Variation (TV, the L1 norm on the first-order finite difference) by White et al. (2017) in \emph{halophot}. However, while \emph{halophot} uses the TV objective function to optimize a set of weights on a fixed set of pixels in an aperture (spatial weighting), we are optimizing a set of aperture corrections that vary in \emph{time}. Objective functions based on first-order finite differences can introduce complex dependencies on the behavior of the underlying signal; for example, for a linear trend with fixed-amplitude Gaussian noise, an L1 or L2 norm on first-order difference increases with decreasing slope, while $\sigma_{p}$ remains constant because of the second-order finite difference used to define it. Under the assumption that within a time series there is a fixed, uncorrelated noise spectrum on top of a well-sampled underlying signal, we adopt $\sigma_{p}$ as our objective function because it is relatively insensitive to the properties of the underlying signal, instead responding to the amplitude of the high-frequency scatter around the underlying signal.

For $k$ apertures, there are $k-1$ aperture corrections to determine and apply. We determine the set of $k-1$ aperture corrections that minimizes $\sigma_p$ using simplex optimization. 

The relatively simple step (v) in the aperture definition process could be replaced by more complex single-aperture optimization approaches, such as those presented in Smith et al. (2016). We investigated watershed segmentation (e.g., Lund et al. 2015) but found that it was not suitable given the rapidly changing, complex topology of the sky flux. An aperture-by-aperture optimization of a set of non-binary aperture weights \'{a} la \emph{halophot} may provide an avenue for future improvement. Regardless of the aperture optimization procedure, the rest of the $k$-means aperture optimization process would remain identical. 

\section{Application to the Titan Lightcurve}

For the Titan ($K_p\sim8.8$ magnitude) dataset of 182 images, we found that $k=5$ apertures was ideal for $k$-means aperture optimization. Using fewer than five apertures resulted in poor photometric performance, while using more apertures resulted in very small incremental improvements that did not merit the additional free parameters introduced. All attempts to optimize a single aperture resulted in poorer performance, with final lightcurve scatter at least two times worse than the nominal 5-aperture solution, and typically showing strong correlation with the background sky flux. The five optimized apertures, the cluster-averaged Titan images for each, and the number of unique images ascribed to each cluster are shown in Figure 1. Note that with $k=5$ clusters, the smallest number of images in this dataset assigned to any cluster is 16. When determining an ideal value for $k$, it is important to consider the the smallest cluster size, as a smaller set of images assigned to any given cluster may introduce a risk of overfitting.

The extracted and merged lightcurves are illustrated in Figures 2 \& 3. The sky lightcurves clearly illustrate one of the chief challenges of this dataset --- the rapidly-varying sky that contributes up to 60\% of the total flux measured in a given aperture. 

The final overall scatter in the lightcurve is $\sigma\sim0.33\%$, while the point-to-point scatter is $\sigma_p \sim 0.19\%$. There is a weak secular trend in the Titan lightcurve. A least-squares line fit to the lightcurve drops by $\sim0.5\%$ over the 3.7-day course of the observations. A Pearson Rank Correlation Coefficient indicates that this trend is significant at $p\leq 0.03$. This secular trend could be due all or in part to either (a) a systematic error in, for example, the estimation of Saturn's scattered light in the Titan aperture, or (b) the 15.945-day rotation of Titan's surface contributing a low-amplitude rotational variation visible through narrow windows of atmospheric transparency in the long-wavelength tail of \emph{Kepler}'s bandpass. Otherwise, the largest apparent feature in the lightcurve occurs at about the mid-time, where a small ($\sim 0.5\%$) apparent brightening occurs over $\sim8$ hours, then fades on the same timescale. The timescale of this brightening is similar to the timescale between \emph{K2} roll-position correction thruster firing events, which could indicate a subtle pointing-related systematic still present in our data. However, a Lomb-Scargle periodogram (Lomb 1976, Scargle 1982) of Titan's lightcurve does not reveal substantial power near the 4 cycle/day peak of thruster firing events (Figure 4), suggesting that the relatively short-period variability in Titan's lightcurve is not in large part attributable to pointing-related systematics. Removing the weak linear correlation between the pointing offsets and Titan's flux (Figure 4) results in negligible change in the final Titan lightcurve.

\subsection{Comparison to Cassini Observations}

During the period of the  \emph{K2} observations (December 3\textsuperscript{rd} and December 7\textsuperscript{th} 2016), there were no observations of Titan from the Cassini Spacecraft. However, clouds were observed prior to (November 14\textsuperscript{th} and November 29\textsuperscript{th} 2016) and after (December 18\textsuperscript{th} and December 30\textsuperscript{th} 2016) the  \emph{K2} observations in Cassini Imaging Science Subsystem (ISS) and Visible and Infrared Mapping Spectrometer (VIMS) measurements, indicating that this was a period of relatively frequent cloud activity (Turtle et al. 2018). In the ISS observations these clouds were reported as ``streaks'' and were not the large (presumably convective) CH\textsubscript{4} cloud systems seen in previous Earth-based observations (e.g., Bouchez \& Brown 2005). These cloud features are low contrast in the broad visible \emph{Kepler} bandpass. An example from March 21, 2017 is given in Figure 5. The broad-band visible image shows extremely low contrast, while the image targeting a narrow NIR band of atmospheric transparency shows a high-contrast cloud streak. The average contrast of the cloud streak with its surroundings is $\sim2\%$ in the broadband image and $\sim10\%$ in the narrowband NIR image. If we adopt $2\%$ as a cloud streak contrast in the  \emph{Kepler} bandpass, a streak would need to subtend $25\%$ of the area of Titan's disk before its contribution to the broadband reflected flux would be comparable to the $\sim 0.5\%$ amplitude features in the  \emph{K2} lightcurve of Titan. For comparison, the cloud streak seen in Figure 5 subtends approximately 8\% of Titan's illuminated disk. Higher contrast broadband atmospheric features would be detectable with commensurately smaller areal coverage, but it is unlikely that cloud streaks similar to those seen by Cassini before and after the  \emph{K2} observations would have been detectable in the lightcurve.


\section{Discussion and Future Applications}

The $k$-means aperture optimization approach may hold further utility in other situations where a moving source is subject to significant background contamination, particularly if the source is saturated. Observations of the planetary satellites (including Titan), planets, and sun-grazing comets by solar monitoring instruments may be one such example. Extracting a lightcurve from the \emph{K2} dataset of Triton or other planetary satellites may also benefit from similar photometric techniques. 

With further effort, $k$-means aperture optimization may provide a useful complement to existing methods for \emph{K2} stellar photometry, particularly in the bright star regime. The pointing drift, periodic thruster firings, and reaction wheel desaturation events that distinguish the two-wheel \emph{K2} mission from the \emph{Kepler} prime mission introduce time variable point-spread functions and substantial focal plane motion even for targets stationary in the sky plane. These introduce substantial systematics in the photometric time series of stationary targets if not carefully accounted for. This challenge has led to the development of a suite of photometry pipelines that mitigate these effects through a variety of means, including decorrelation against spacecraft pointing with  piecewise-linear models (e.g., Vanderburg \& Johnson 2014) or via Gaussian process models (Aigrain et al. 2015, 2016), or by PLD (Luger et al. 2018). $K$-means aperture optimization represents a novel means of mitigating image-plane effects like those introduced in \emph{K2}, so we performed a test to compare the performance of the basic algorithm on an $K_p=9.2$ magnitude eclipsing binary star in \emph{K2} Campaign 0 data (EPIC 202063160); using the same prescriptions as with Titan, but with $k=30$ apertures over 1,641 images rather than $k=5$ over 182 images. We illustrate the results in Figure 6. \emph{k}-Means aperture optimization delivers a lightcurve with approximately twice the scatter of EVEREST 2.0 (Luger et al. 2018). With further refinements made to the aperture optimization step, it is possible that $k$-means aperture optimization will be competitive with the state of the art for saturated stars in $K2$, providing an alternative pathway to precision photometry under conditions where other methods are prohibited.

Long-term monitoring of Titan's variability from a stable space-based platform holds great promise for building a comprehensive picture of the power spectrum of variability in its complex atmosphere. While the  \emph{K2} lightcurve of Titan shows no compelling evidence for variability that is concretely attributable to the atmosphere, observations of similar precision in more appropriate spectral bandpasses would be able to track far smaller atmospheric changes than those that could have been revealed by the current  \emph{K2} lightcurve.

\section{Acknowledgements}

The analysis in this work was supported by the NASA grant 15-K2-GO4\_2-100. This research has made use of NASA's Astrophysics Data System Bibliographic Services. This research made use of numerous community-developed software packages, including Astropy (Robitaille et al. 2013, Price-Whelan et al. 2018), Numpy (Oliphant 2006), Matplotlib (Hunter 2007), SciPy (Jones et al. 2001), scikit-learn (Pedregosa et al. 2011), and scikit-image (van der Walt et al. 2014). We wish to thank the referee for a very thorough and constructive review.


\begin{figure}[hb]
\includegraphics[width=15.5cm]{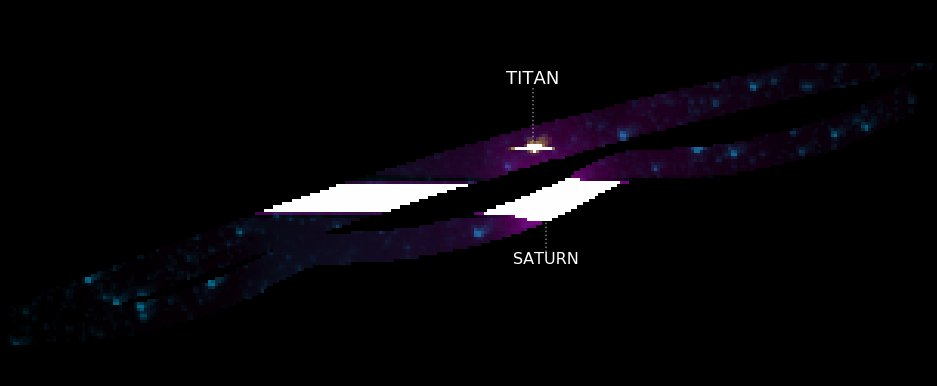}
\includegraphics[width=15.5cm]{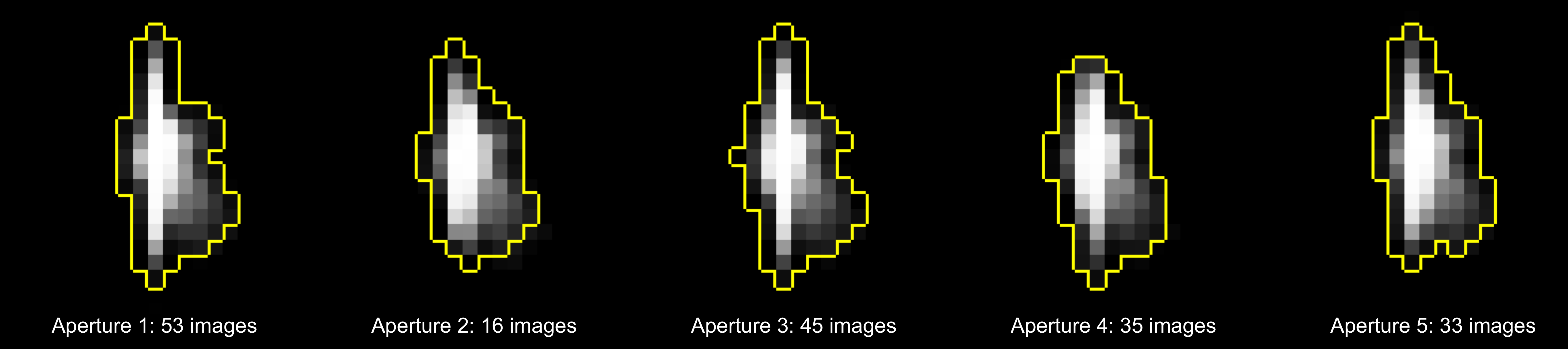}
\caption{Top panel: Individual frame from the \emph{K2} Titan ($K_p\sim8.8$ magnitude) time series. Saturated pixels shown in white; DN associated with Saturn scattered light shown in magenta, DN associated with Titan PSF shown in yellow, and DN associated with background stars shown in cyan. Frame collected UTC 2016-Dec-06 11:49:48.5. Bottom panel: The five optimized apertures (yellow) applied to the  \emph{K2} Titan data, determined using $k$-means aperture optimization. Apertures are overlaid on the cluster-average image of Titan, showing the variation in the PSF shape and saturation charge bleeds due to the average sub-pixel positioning of Titan in each identified cluster.}
\label{fig1}
\end{figure}

\begin{figure}
\includegraphics{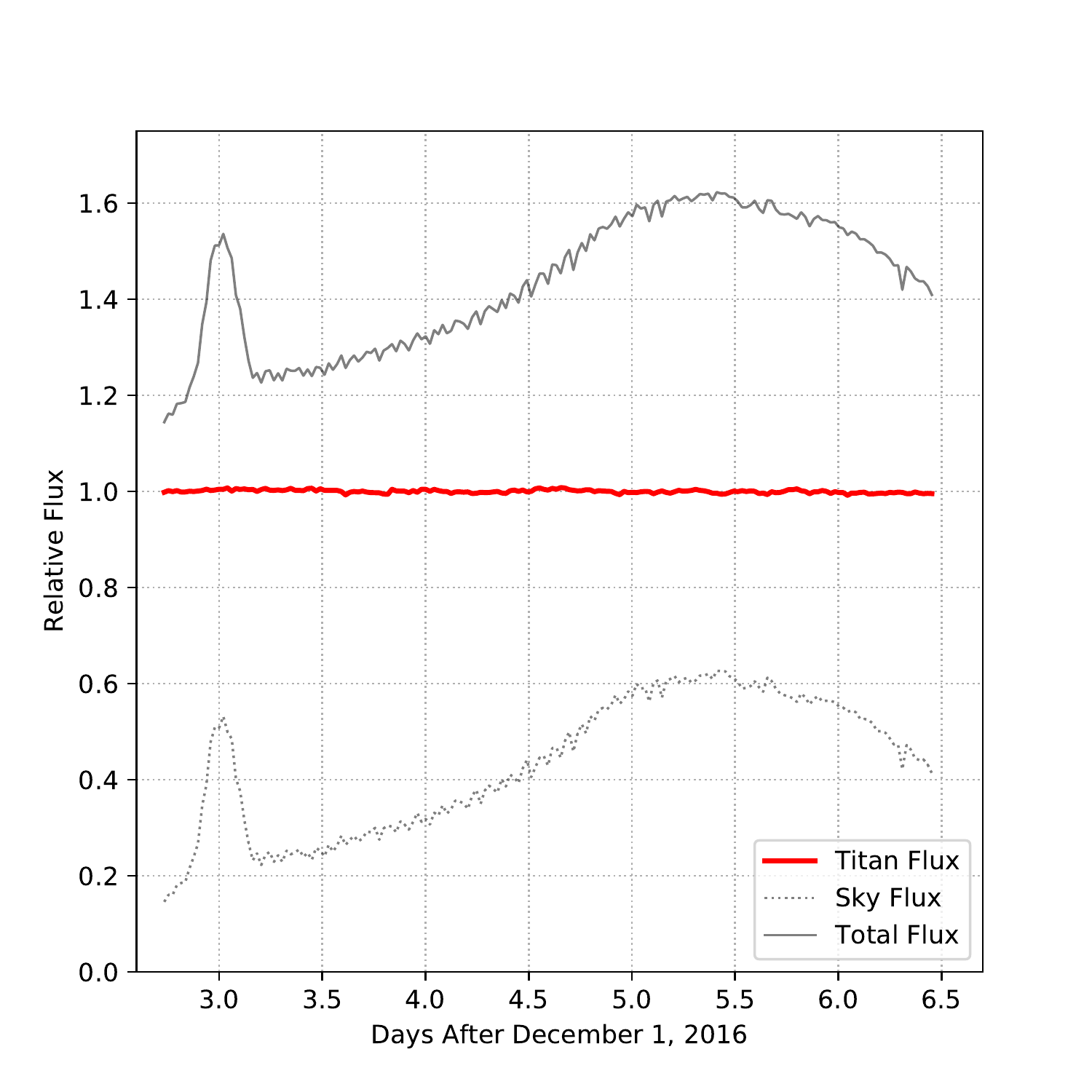}
\caption{ \emph{K2} lightcurve of Titan (red) after $k$-means aperture optimization, showing estimated background ``sky'' flux in the aperture (gray dash) and total measured aperture flux (gray solid). The rapid changes in sky flux is due to Titan's motion through the scattered light of Saturn, as well as rapid temporal changes in that scattered light as Saturn moved across a chip boundary. All lightcurves are shown relative to the mean of Titan's flux over the observation.}
\label{fig2}
\end{figure}

\begin{figure}
\includegraphics{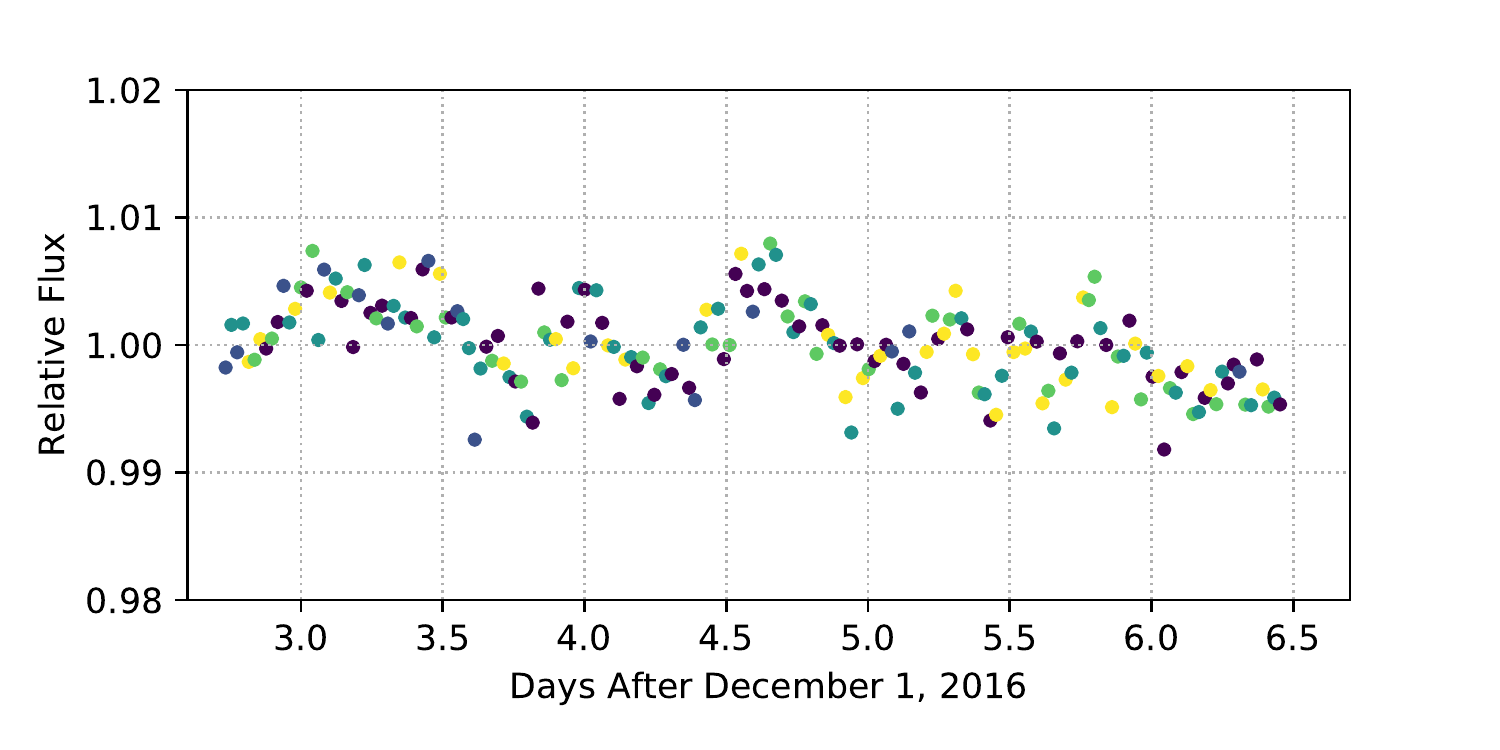}
\caption{ \emph{K2} lightcurve of Titan, scaled to illustrate final scatter. The lightcurve is shown relative to the mean of Titan's flux over the observation. Point color indicates which of the five optimized apertures was applied to that point. Apertures are well distributed in time. Point-to-point scatter estimated to be 0.19\%.}
\label{fig3}
\end{figure}

\begin{figure}
\includegraphics[width=15.5cm]{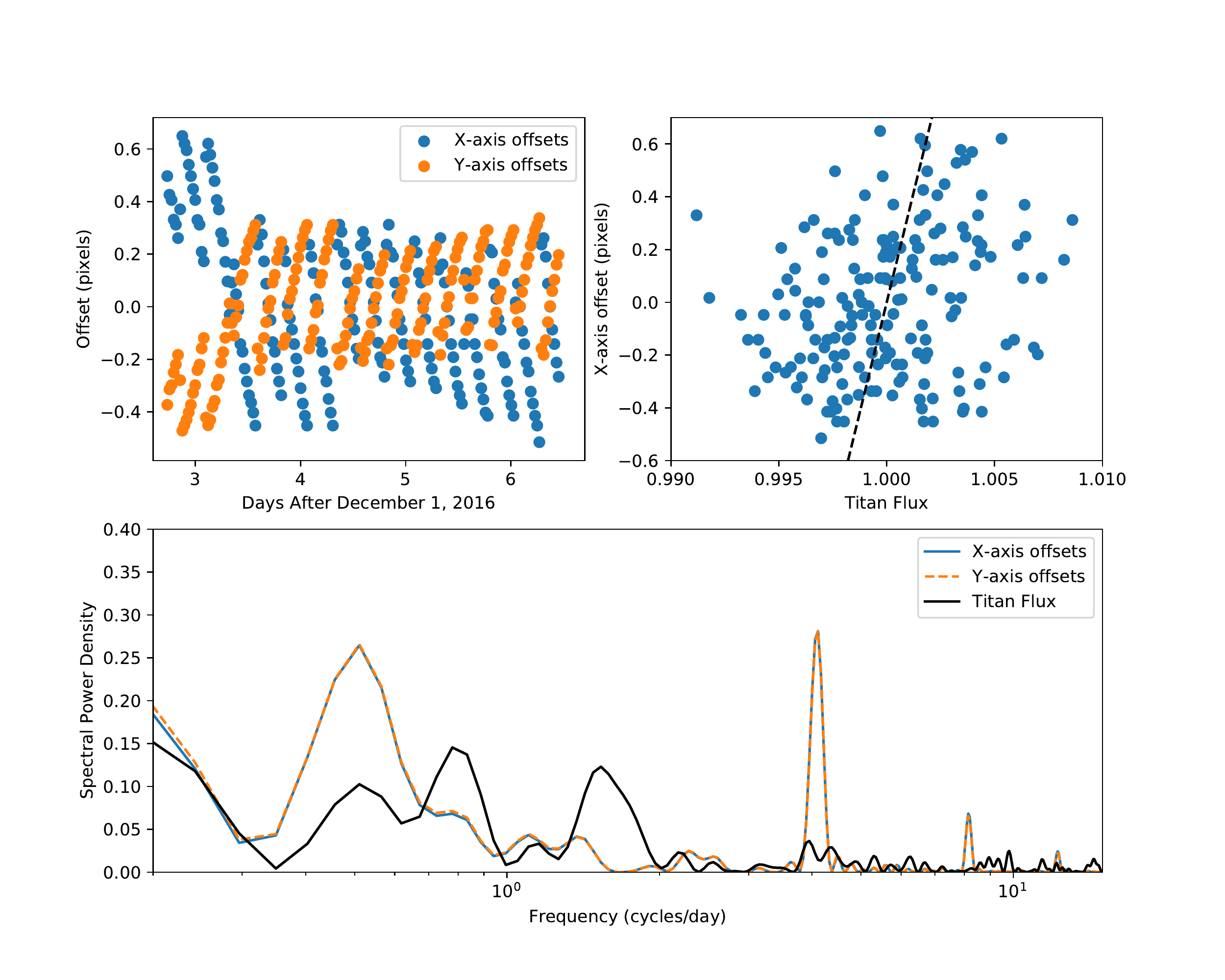}
\caption{Top left: WCS offsets measured per frame of the Titan dataset. The X and Y offsets are strongly anti-correlated. Top right: Final Titan lightcurve flux versus X-axis WCS offset. A slight residual linear correlation is present, illustrated by the black dashed line. Removing this linear trend results in negligible change in the lightcurve. Bottom panel: Lomb-Scargle periodograms of the X and Y axis WCS offsets contrasted against a Lomb-Scargle periodogram of Titan's flux. The 4 cycle/day peak in the WCS offsets is due to the periodic thruster firing events correcting spacecraft roll angle. This frequency has very little power in the Titan periodogram, which instead shows peaks at 0.8 and 1.75 cycles/day.}
\label{fig4}
\end{figure}

\begin{figure}
\centering
\includegraphics[width=6cm]{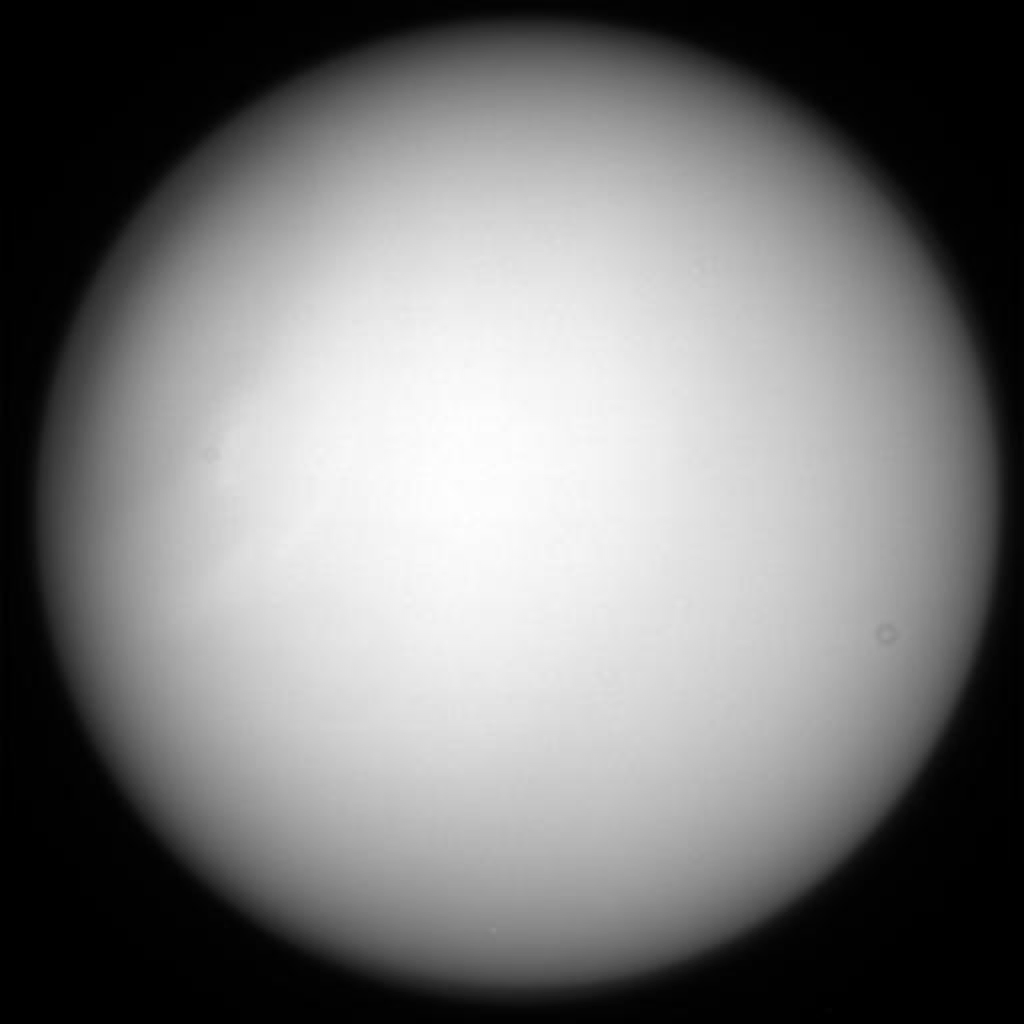}
\includegraphics[width=6cm]{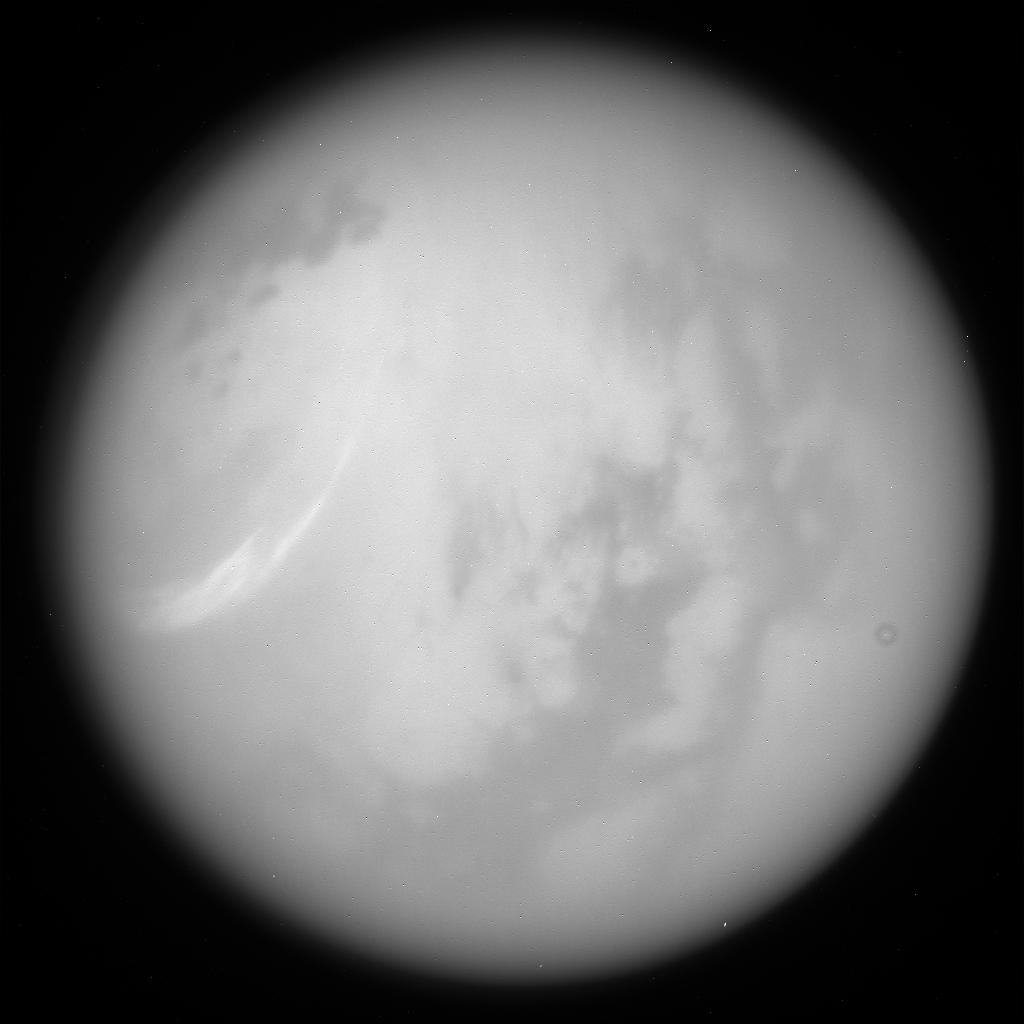}
\caption{Example of a cloud streak seen by the Cassini ISS on March 21, 2017 in the GRN (left) and CB3 (right) filters (images N00278458 and N00409517, respectively). The broad-band GRN image shows very low contrast, while the narrowband NIR CB3 image shows the bright cloud streak (in the 9-o'clock position) at much higher contrast. Dark surface features are also visible in the CB3 image. All  \emph{K2} observations are taken in a very broad optical bandpass which will result in contrast similar to the left panel for tropospheric clouds.}
\label{fig5}
\end{figure}

\begin{figure}
\centering
\includegraphics[width=15.5cm]{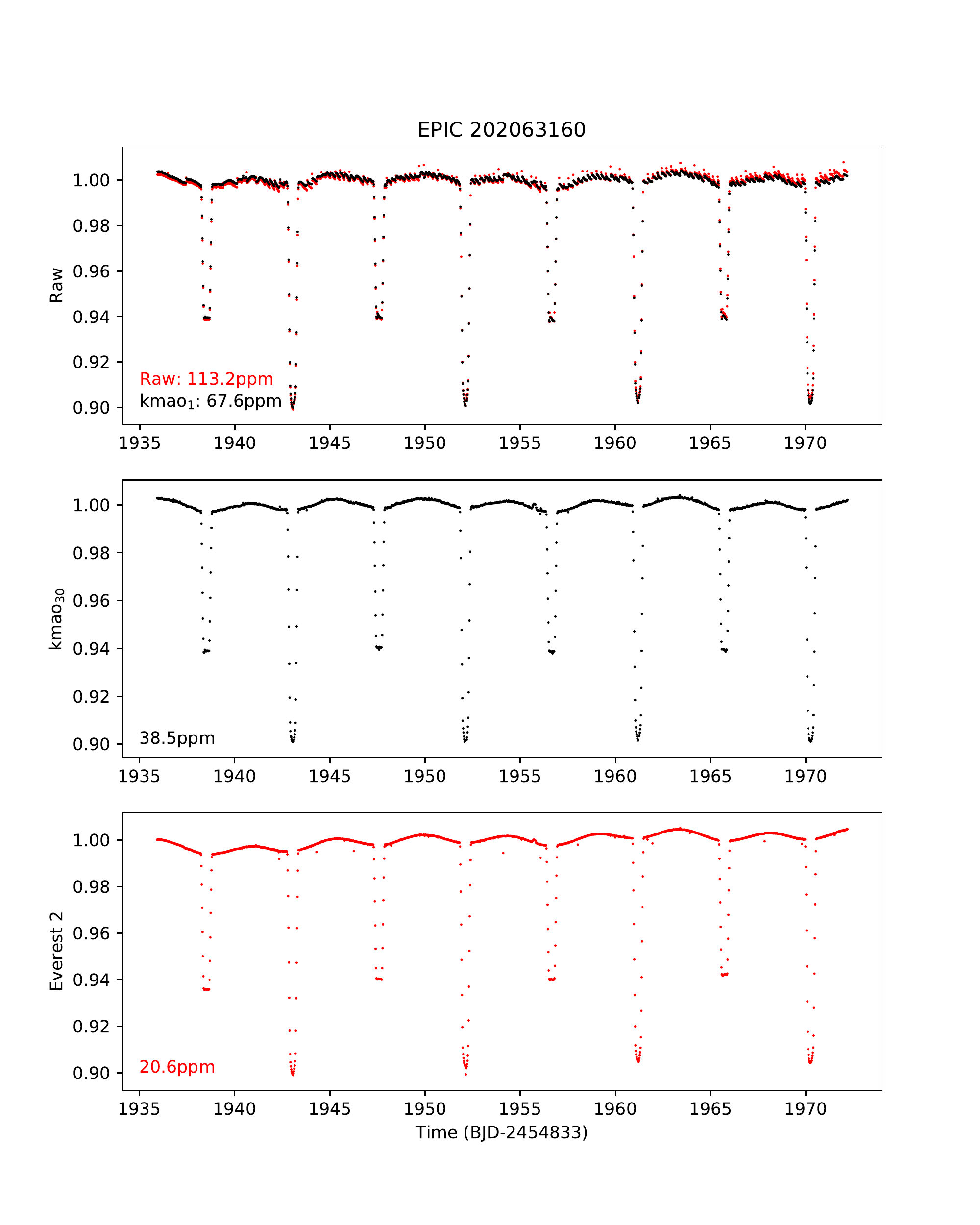}
\caption{Lightcurve of \emph{K2} Campaign 0 eclipsing binary EPIC 202063160 reduced via $k$-means aperture optimization (single-aperture and 30 aperture versions), and comparison to raw and EVEREST 2.0 lightcurves. Quoted scatter is an approximation of the LC 6-hour scatter, $\sigma_{p}/\sqrt{12}$.}
\label{fig6}
\end{figure}

\end{document}